%% file: main.tex
\RequirePackage{amsmath}
\def\year{2021}\relax
\documentclass[letterpaper]{article} 

\input{macro}

\usepackage{anyfontsize}

\usepackage{graphicx} 

\usepackage{svg}
\usepackage{wrapfig}
\usepackage{xcolor}
\usepackage{booktabs}
\usepackage{multirow}

\usepackage[T1]{fontenc} 
\usepackage[utf8x]{inputenc}

\usepackage[export]{adjustbox}

\newcommand{\richb}[1]{\textbf{\color{red} richb: #1}} 

\usepackage{aaai21}  
\usepackage{times}  
\usepackage{helvet} 
\usepackage{courier}  
\usepackage[hyphens]{url}  
\usepackage{graphicx} 
\usepackage{natbib}  
\usepackage{caption} 
\title{Educational Question Mining At Scale: Prediction, Analysis and Personalization}

\author {
        Zichao Wang,\textsuperscript{\rm 1}\thanks{Work done during internship at Microsoft Research Cambridge.}
        Sebastian Tschiatschek,\textsuperscript{\rm 2}\thanks{Work done at Microsoft Research Cambridge.}
        Simon Woodhead,\textsuperscript{\rm 3}
        Jos\'e Miguel Hern\'andez-Lobato,\textsuperscript{\rm 5$\dagger$}
        Simon Peyton Jones,\textsuperscript{\rm 4}
        Richard G. Baraniuk,\textsuperscript{\rm 1}
        Cheng Zhang\textsuperscript{\rm 4}\\
}
\affiliations {
    \textsuperscript{\rm 1}Rice University
    \hspace{5pt}
    \textsuperscript{\rm 2}University of Vienna 
    \hspace{5pt}
    \textsuperscript{\rm 3}Eedi
    \hspace{5pt}
    \textsuperscript{\rm 4}Microsoft Research Cambridge
    \hspace{5pt}
    \textsuperscript{\rm 5}University of Cambridge
    \\[3pt]
    {\fontsize{10}{10}{\{jzwang, richb\}@rice.edu}\hspace{15pt} { sebastian.tschiatschek@univie.ac.at} \hspace{15pt} { jmh233@cam.ac.uk}\\
    { simon.woodhead@eedi.co.uk}\hspace{15pt} { \{simonpj, Cheng.Zhang\}@microsoft.com}}
}

\begin{document}

\maketitle

\input{abstract}
\input{intro}

\input{dataset}
\input{method}

\input{experiment}

\input{related}

\input{discussion}

\section{Acknowledgements}
We thank the anonymous reviewers for their constructive feedback. ZW and RGB are supported by NSF grants 1842378 and 1937134 and by ONR grant N0014-20-1-2534.


{
\fontsize{10pt}{10.0pt}
\bibliography{bib.bib}
}
\end{document}

%% file: macro.tex
\usepackage{amsmath,amsfonts,bm, bbm}

\def\eqref#1{equation~\ref{#1}}

\def\1{\bm{1}}

\def\rve{{\mathbf{e}}}

\def\rvz{{\mathbf{z}}}

\def\ve{{\bm{e}}}

\def\vs{{\bm{s}}}

\def\vx{{\bm{x}}}

\def\vz{{\bm{z}}}

\def\mX{{\bm{X}}}

\DeclareMathAlphabet{\mathsfit}{\encodingdefault}{\sfdefault}{m}{sl}
\SetMathAlphabet{\mathsfit}{bold}{\encodingdefault}{\sfdefault}{bx}{n}



%% file: abstract.tex
\begin{abstract}
Online education platforms enable teachers to share a large number of educational resources such as questions to form exercises and quizzes for students. With large volumes of available questions, it is important to have an automated way 
to quantify their properties and intelligently select them for students, enabling effective and personalized learning experiences. In this work, we propose a framework for mining insights from educational questions at scale. We utilize the state-of-the-art Bayesian deep learning method, in particular partial variational auto-encoders (p-VAE), to analyze real students' answers to a large collection of questions. Based on p-VAE, we propose two novel metrics that quantify question quality and difficulty, respectively, and a personalized strategy to adaptively select questions for students. We apply our proposed framework to a real-world dataset with tens of thousands of questions and tens of millions of answers from an online education platform. 
Our framework not only demonstrates promising results in terms of statistical metrics but also obtains highly consistent results with domain experts' evaluation. 
\end{abstract}

%% file: intro.tex
\section{Introduction}
Online education platforms are transforming education by 
democratizing access to high-quality educational resources and personalizing learning experiences. 
A central instrument in today's online education scenarios is assessment questions (referred to as ``questions'' henceforth), which help teachers evaluate the students' abilities and help students reinforce the knowledge they are learning~\cite{doi:10.1177/0963721412443552, Karpicke966, Patricia2000, Kovacs:2016:EIQ:2876034.2876041, Koedinger:2015:LSS:2724660.2724681}. 
Questions are particularly important in {\it online} education, because teachers have more limited interactions with students; students answer records to questions serve as one of the few ways for teachers to interact with and understand their students~\cite{twigg2003models}. 
With the world reeling from the impact of the Covid-19 pandemic, there has been a rapid and massive increase in the number of students learning online~\cite{bao2020covid,sandars2020twelve,verawardina2020reviewing}.
Questions as assessment and learning instruments have thus become even more prominent. 

A key challenge to best utilize these questions is how to choose the most suitable ones for students. Only high quality and suitably difficult questions are beneficial for learning~\cite{blanchette2001questions}; as a result, understanding these two properties of questions help guide the choice of questions. While in traditional classroom settings, manually examining and selecting questions to attend to the learning status of each individual student remains the best practice, this labor- and time-intensive procedure clearly does not scale to large-scale online education scenarios in which the number of questions and students can be massive.  
We thus desire an efficient tool for question analysis at scale and capable of computing question analytics including quality and difficulty and automatically selecting questions for students. The analytics will serve as side information that helps teachers and students select appropriate questions for their pedagogical and educational needs. The automatic question selection process enables personalized learning and adaptive testing experience when the number of students greatly exceeds teachers' capacities~\cite{chang2009nonlinear,yan2016computerized}. Both analytics and personalization are important to realize the values of massive questions in online education scenarios.

To address the above challenges, we aim to develop an AI solution for large-scale online educational question mining, 
providing both insights for question quality and difficulty and strategies for choosing questions for each student. 
This task involves 3
challenges. First,  
the number of questions, students, and answer records is extremely large and processing such data may be computationally expensive.
Second, online educational data is highly incomplete because each student can only answer a small fraction of all available questions. 
Third, we need to design quantitative metrics and strategies to accurately identify question quality and to adaptively choose questions for each student. 
Overall, we need a solution that is efficient, handles highly sparse data, and automatically acquires educationally meaningful and actionable insights about questions.

\subsection{Contributions}
In this work, we collect a large real-world online educational dataset in
the form of 
students' answers to multiple-choice questions and develop a machine learning framework to quantify question quality and difficulty and provide personalized question selection strategy.
We briefly summarize our framework below.

\begin{itemize}
\item We leverage the partial variational auto-encoder (p-VAE)~\cite{ma2018partial,eddi} to efficiently handle the highly sparse educational data at a large scale. p-VAE models existing students' answers and predicts the potential answer to unseen questions in a probabilistic manner.

\item We propose a novel information-theoretic metric to quantify the quality of each question based on p-VAE.  
We also define a metric to quantify question difficulty.

\item We propose a novel information-theoretic strategy to sequentially select questions for each student. Our strategy chooses a sequence of questions which best identifies the student's learning status. 

\item We experimentally validate 
our framework on a new, large educational dataset and demonstrate state-of-the-art performances of our framework over various baselines.

\end{itemize}

%% file: dataset.tex
\section{Dataset}
\label{sec:cohort}

We analyze data from Eedi,\footnote{\url{https://eedi.com/projects/neurips-education-challenge}} a renowned real-world online education platform used by over 100,000 thousand students and 25,000 teachers in over 16,000 schools. Eedi offers crowd-sourced, multiple-choice diagnostic questions to students from primary to high school (roughly between 7 and 18 years old).
Each question has 4 answer choices and only one of them is correct.
Currently, the platform focuses mainly on math questions. Figure~\ref{fig:example_question} shows an example question from the platform. 
We use data collected from the 2018 -- 2019 school year. 

We organize the data in a matrix form where each row represents a student and each column represents a question. Each entry contains a number that represents whether the student has answered a question correctly (i.e., 0 represents the wrong answer and 1 represents the correct answer). Each student has only answered a tiny fraction of all questions and hence the matrix is extremely sparse. We thus removed questions that contain less than 50 answers and students who have answered less than 50 questions. Besides, when a student has multiple answer records to the same question, we keep the latest answer record. 

The above preprocessing steps lead to a final data matrix that consists of more than 17 million students' answer records with 
123,889 students (rows) and 27,613 questions (columns), making it one of the largest educational datasets to date compared to a number of existing ones~\cite{assistment, bridge, Algebra}. 
Additionally, each question is linked to one or more topics that describe the skills~\cite{vanlehn1988student, chrysafiadi2013student} that the question intends to assess. We have open-sourced this dataset~\cite{wang2020diagnostic} and it is available at { \url{https://eedi.com/projects/neurips-education-challenge}}.

%% file: method.tex
\section{Method}

The extreme sparsity and massive quantity 
of our dataset
brings challenges to analyze both the question quality because each student only answers a small fraction of potentially non-overlapping questions. To gain insights into such real-world educational data, we first need a model that predicts the missing data with uncertainty estimation. The missing data in our case is students' answers to unseen questions. With such a model, we can design different metrics to quantify question quality and difficulty. 

The first step is formulated as the following probabilistic missing data imputation (matrix completion) problem. We have a data matrix $\mX$ of size $N$ by $M$, where N is the total number of students and M is the total number of questions. Each entry $x_{ij}$ is binary which indicates whether student $i$ has answered question $j$ correctly.\footnote{$x_{ij}$ can also be categorical which is the answer in a multiple choice question that a student selects.}
The data matrix is only partially observed; we denote the observed part of the data matrix as $\mX_{O}$. Then, we would then like to accurately predict the missing entries in a {\it probabilistic} manner which enables the design of various metrics for question quality, difficulty, and selection strategy. 
Thus, we use the partial variational auto-encoder (p-VAE)~\cite{ma2018partial}, which is the state-of-the-art method for the above imputation task.

The second step is to quantify question quality and difficulty and adaptively select questions for students.
Specifically, we define a difficulty measure using the full data matrix completed by p-VAE. We quantify question quality by measuring the value of information that each question carries using an information theoretical metric. Using Similar ideas, we sequentially select questions for each student based on a notion of information gain. Such ``information'' computation is made possible and efficient by our utilization of the p-VAE in the first step above.
We present these two steps in detail in the remainder of this section.

\begin{figure}[t]
    \centering
    \includegraphics[width=0.55\linewidth]{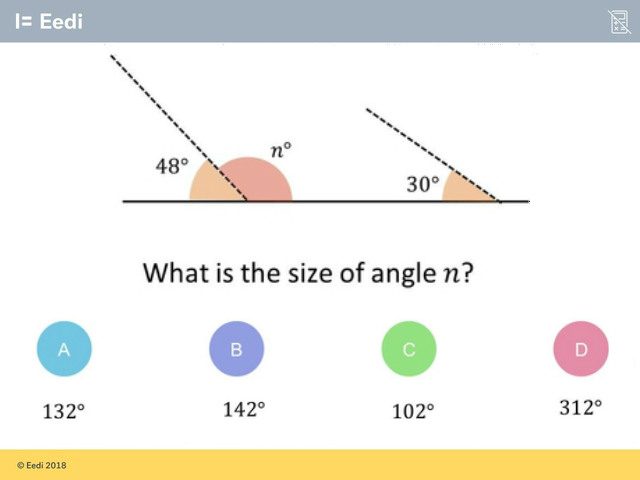}
    \caption{An example question from our new dataset.}
    \label{fig:example_question}
    \vspace{-10pt}
\end{figure}

\subsection{Partial Variational Auto-encoder (p-VAE) for Student Answer Prediction}

p-VAE~\cite{eddi} is a deep latent variable model that extends traditional VAEs~\cite{kingma2013auto,Rezende2014StochasticBA,8588399} to handle missing data as in such education applications. 
VAEs assume that the responses $\vx_{i}$ of student $i$  is generated from a latent variable $\rvz_i$:
\begin{align}
    p_\theta(\mX) = \prod_{i=1}^N p_\theta(\vx_{i}) &= \prod_{i=1}^N\int p_\theta(\vx_{i} | \rvz_i)\, p_\theta(\rvz_i)\,d\rvz_i \nonumber\\
    &= \prod_{i=1}^N \int \prod_{j=1}^Mp_\theta(x_{ij} | \rvz_i)\,p_\theta(\rvz_i)\,d\rvz_i\,,\nonumber
\end{align}
where 
$x_{ij}$ is the $i$-th student's answer to the $j$-th question. We use a deep neural network for the generative model $p_\theta(x_{ij} | \rvz_i)$ because of its expressive power. Of course, $\vx_{i}$ contains missing entries because each student $i$ only answers a small fraction of all questions.
Unfortunately, VAEs can only model fully observed data. To model partially observed data, p-VAE extends traditional VAEs by exploiting the fact that, given $\vz_i$, $\vx_i$ is fully factorized. Thus, the unobserved data entries can be predicted given the inferred $\rvz_i$'s. Concretely, p-VAE optimizes the following {\it partial} evidence lower bound (ELBO):
\begin{align} \label{eq:pVAE}
& \log p(\mX_O) \geq \log p(\mX_O) - D_{\textrm{KL}}(q(\mathbf{z}|\mX_O) \| p(\mathbf{z}|\mX_O)) \nonumber \\ \nonumber
& = \mathbb{E}_{\mathbf{z} \sim q(\mathbf{z}|\mX_O)}[  \log p(\mX_O|\mathbf{z}) + \log p(\mathbf{z}) - \log q(\mathbf{z}|\mX_O)]\\ \nonumber
&\equiv \mathcal{L}_{\rm partial}\,,
\end{align}
which is in the same form as the ELBO for VAE but only over the observed part of the data. 
Because $\mX$ is binary, we use Bernoulli distribution as the likelihood function. However, we can also choose to use students' actual answers (A, B, C, or D) as $\mX$ where the likelihood function becomes categorical. Investigation of categorical data format is left for future work.

\begin{figure}
    \centering
    \includegraphics[width=0.75\linewidth]{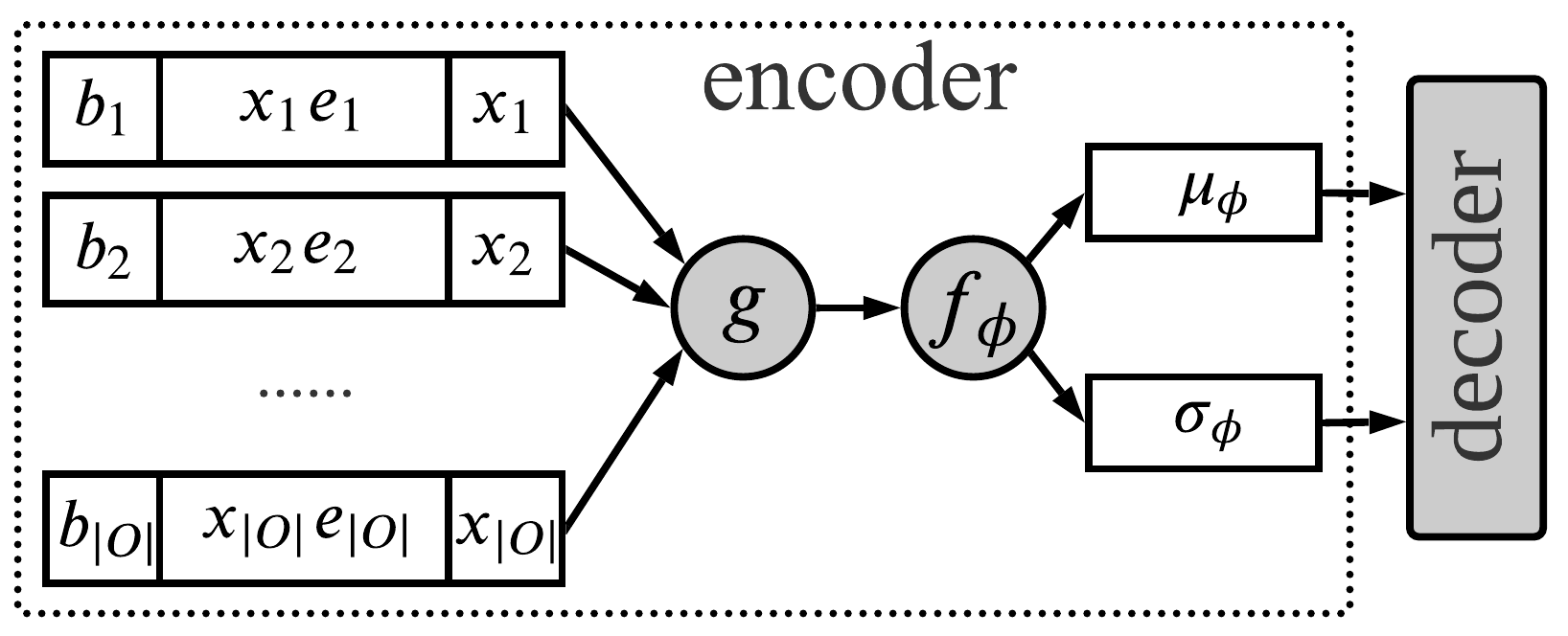}
    \vspace{-3pt}
    \caption{Illustration of the p-VAE model architecture.}
    \label{fig:pvae}
\end{figure}

The challenge is to approximate the posterior of $\rvz_i$'s using a partial observed data vector. p-VAE uses a set-based inference network $q_\phi(\rvz_i | \vx_{O_i})$, where $\vx_{O_i}$ is the observed subset of answers for student $i$ \cite{zaheer2017deep,qi2017pointnet}. 
$q_\phi(\rvz_i | \vx_{Oi})$ is assumed to be Gaussian; Concretely, the mean and variance of the posterior of the latent variable is inferenced as
\begin{align}
    [\mu_\phi(\vx_{O}), \sigma_\phi(\vx_{O})] &= f_\phi(g(\vs_1, ...\,, \vs_{ij}, ...\,, \vs_{|O|}))\,,
\end{align}
where we have dropped the student index $i$ for notation simplicity.
$\vs_{ij}$ is the observed answer value augmented by its location embedding, which is learned;  $g(\cdot)$ is a permutation invariant transformation such as summation which outputs a fix sized vector; and $f_\phi: \mathbb{R}^M\rightarrow\mathbb{R}^K$ is a regular feedforward neural network. 
In this paper, we set $\vs_{ij} = [x_{ij}, x_{ij}\ve_j, b_j]$
where 
Figure~\ref{fig:pvae} illustrates the network architecture of p-VAE.

Note that, in p-VAE, some parameters have natural interpretations. For example, the per-question parameters $[\rve_j, b_j]$ can be collectively interpreted as a {\it question embedding} for each question $j$. The per-student latent parameter $\rvz_i$ can be interpreted as a {\it student embedding} for each student $i$. 
Furthermore, p-VAE quantifies the uncertainty of the student embedding $\rvz_i$ in the form of an (approximate) posterior distribution. This enables us to define information-theoretic metrics and strategies for question difficulty, question quality, and question selection, which we describe next.

\subsection{Question Difficulty Quantification}
For a group of questions answered by the same group of students, the difficulty level of the questions can be quantified by the incorrect rate of all students' answers. However, for real-world online education data, every question is answered only by a small fraction of students and by different subsets of students with different educational backgrounds. Thus, directly comparing the difficulty levels of the question from observational data is not accurate because an easy question, which may be answered by only less-skilled students, may be shown to be difficult if only observational data is used. 
Thanks to p-VAE, we can predict whether a student can correctly answer an unseen question. We achieves this by first predicting students' responses to all unanswered questions and then defining the difficulty level of question $j$ as $\sum_i^N \frac{p(x_{ij}=0)}{N} \approx \frac{1}{N}\sum_i^N \hat{x}_{ij}$ where $\hat{x}_{ij}$ denotes if the student has answered the question correctly. 
Higher value implies that more students are predicted to answer this question correctly and that this is an easier question.

\subsection{Question Quality Quantification}
Working closely with education experts, we found that high-quality questions are considered to be those that best differentiate student abilities. When a question is simple, almost all students will answer it correctly. When a question is badly formulated, all students will provide incorrect answers or random guesses. In any of these cases, the question neither helps the teacher gain insights about the students' abilities nor helps students learn well. Thus, high-quality questions are the ones that can differentiate the students' abilities.

We thus formulate the following information theoretic objective to quantify the quality of question $j$: 
\begin{align}
    R(j) &= \mathbb{E}_{x_{ij}\sim p_\theta(x_{ij})}\Big[{\rm D_{KL}}[p_\phi(\rvz_i | x_{ij}) \,||\, p(\rvz)]\Big]\,,
    \label{eq:question-quality-form-1} \\
    &\approx\frac{1}{S}\sum_{i=1}^S{\rm D_{KL}}\left[q_\phi(\rvz | x_{ij}) | p(\rvz)\right]\,,
    \label{eq:question-quality-form-2}
\end{align}
where we have used Monte Carlo integration and replaced $p_\phi(\rvz_i|x_{ij})$ with $q_\phi(\rvz_i|x_{ij})$ for practical and efficient computation~\cite{eddi, gong2019icebreaker}.
$j$ is the question index, $x_{ij}$ is the $i$-th student's answer to the $j$-th question, which can be either binary indicating whether the student has answered it correctly or categorical which is the student's answer choice for this question. $\rvz$ is the latent embedding of students. The $i$-th student ability can be determined by the student's possible performance on all questions, which can be inferred from $\rvz_i$. 

This objective measures the information gain of estimating the student ability by conditioning on the answer to question $j$. When $R$ is large, the question is more informative on differentiating the student ability reflected in the student embedding $\rvz$, and thus it is considered as high-quality.

\subsection{Personalized Question Selection}
In practical online education scenarios, it is of great interest to adaptively select a small sequence of questions for students. Appropriately choosing these questions allows effective and efficient evaluation of students' abilities at scale. 

We 
formulate the problem of personalized question selection as a Bayesian experiment design problem in an information theoretic manner. Specifically, we are interested in selecting a sequence of questions that is most informative in revealing the student's current state of learning. Inspired by~\cite{eddi, gong2019icebreaker}, we formulate the following selection strategy which is similar to Eq.~\ref{eq:question-quality-form-2}:
\begin{align}
    R_i(j, \vx_{O}) =\mathbb{E}_{x_{j}\sim p(x_j|\vx_{O})} D_{\rm KL}[p(\rvz_j|\vx_O, x_{j}) || p(\rvz_j | \vx_{O})]\,. \label{eq:rec-1}
\end{align}
Intuitively, this selection strategy selects a question $j$ that provides the most information, defined in KL divergence, on the student knowledge state summarized in $\rvz_i$. This objective is different from the one used in~\cite{eddi} as we do not have a particular target variable.
Note that we have omitted the student index subscript $i$ in~Eq.~\ref{eq:rec-1} for succinctness. 
Similar to the previous quality metric computation, we use Monte Carlo integration to approximate Eq.~\ref{eq:rec-1}. 

The above selection strategy enables selecting a sequence of personalized questions for each student in the following manner (which is different from Eq.~\ref{eq:question-quality-form-2} which can only select a single question). First, we initialize $\vx_O = \emptyset$. Then, we compute the information reward according to Eq.~\ref{eq:rec-1} for each question and select question $j$ outside the prediction targets (i.e., $x_j\notin\vx_\psi$) with the maximum $\widehat{R}$ as the next one for student $i$. Finally, we set $\vx_O \leftarrow \vx_O \bigcup x_j$ and repeat the previous steps until we reach the desired number of selected questions.

%% file: experiment.tex
\section{Experiments}

In this section, we demonstrate the applicability of our framework on the real-world educational dataset that we have introduced in the dataset section for student answer prediction, question difficulty, and quality quantification, and personalized question selection.

\subsection{Student Answer Prediction}

\paragraph{Setup}  We split the students (rows of the data matrix) into the train, validation, and test sets with an 80:10:10 ratio. Therefore, students that are in the test set are never seen in the training set.
We train the model on the train set for 50 epochs using Adam optimizer~\cite{kingma2014adam} with a learning rate of 0.001. 
We train p-VAE on binary students' answer records (correct or incorrect answers).
To evaluate imputation performance, we supply the trained p-VAE model a subset of the test set as input and compute the model's prediction accuracy and mean absolute error (MAE) on the rest of the test set.

\begin{table}[t]
\centering
{\small
\begin{tabular}{@{}lcc@{}}
\toprule
{\bf Method} & {\bf Accuracy $\uparrow$} & {\bf Mean Abs. Err. $\downarrow$} \\ \midrule
Random & 0.534 & 0.471 \\
IRT & 0.735 & 0.359 \\
SVD & 0.734 & 0.358 \\
SVD++ & 0.737 & 0.352 \\
Co-clustering & 0.731 & 0.351 \\
NMF & 0.737 & 0.382 \\
\textbf{p-VAE (ours)} & {\bf 0.739} & {\bf 0.343} \\ \bottomrule
\end{tabular}}
\caption{Imputation performances of various methods. p-VAE remains a very strong competitor and slightly outperforms all baselines that we consider. 
}
\label{tab:accuracy}
\end{table}

\paragraph{Results} Table~\ref{tab:accuracy} shows the accuracy of p-VAE trained on binary answer records comparing to various baselines, 
including random imputation, the Rasch model~\cite{rasch1960studies}, SVD, SVD++~\cite{10.1145/1401890.1401944}, Co-clustering~\cite{10.1109/ICDM.2005.14} and Negative Matrix Factorization (NMF)~\cite{6748996}. Note that SVD can also be interpreted as a multivariate Rasch model~\cite{chalmers2012mirt} which is another classic method in educational data mining. We did not compare with vanilla VAEs~\cite{kingma2013auto} because, as mentioned previously, they cannot handle an incomplete input data matrix.

We can observe that p-VAE achieves state-of-the-art performance on the dataset, slightly outperforming all baselines. In addition, p-VAE is the only method that not only accurately predicts students' responses but also computes a posterior distribution over student embeddings which enables the computation of question quality and personalized question selection. The remaining baselines either do not perform as well as p-VAE or do not compute any uncertainty information. There exist Bayesian versions of some baselines, such as Bayesian IRT~\cite{doi:10.3102/10769986007003175} or Bayesian sparse factor analysis~\cite{Lan:2014:TLC:2623330.2623631} models. Unfortunately, due to the high computational cost, these traditional Bayesian models do not scale to datasets as big as the one that we consider in this work. Thanks to the amortized inference, p-VAE is capable of efficient Bayesian inference at scale.

\paragraph{Implications} Accurately predicting students' answers and estimate uncertainty lies at the core of educational data mining and adaptive testing. p-VAE achieves both with state-of-the-art performances which is a necessary prerequisite to the computation involved in the remaining experiments.

\subsection{Question Difficulty Quantification}
\paragraph{Setup} With the complete data matrix imputed by p-VAE, we compute question difficulty by taking the average of all students' answers including observed and predicted answers.
We resort to human evaluation to compare our framework's rankings. We ask the evaluator to provide a full difficulty ranking for all topics. We then compute the difficulty ranking using our framework and compute the Spearman correlation coefficient as a measure of the level of agreement between the model's and the expert's difficulty rankings. 

\begin{table}[]
\centering
{\small
\begin{tabular}{@{}lc@{}}
\toprule
{\bf Method} & {\bf Spearman Correlation} \\\midrule
random ordering &  0.060  \\
majority imputation  & 0.076  \\
using observation  & 0.395  \\
{\bf p-VAE imputation}  & {\bf 0.738} \\ \bottomrule
\end{tabular}}
\vspace{-5pt}
\caption{Spearman Correlation coefficients for question topic difficulty rankings between human expert and model prediction.}
\label{tab:spearmanr-scheme}
\end{table}

\begin{table}
\centering
{\small
\begin{tabular}{@{}lccccc@{}}
\toprule
{\bf Method} & {\bf T1} & {\bf T2} & {\bf T3} & {\bf T4} & {\bf T5} \\ \midrule
Random & 0.4 & 0.56 & 0.44 & 0.44 & 0.44 \\
Entropy & 0.68 & 0.64 & 0.64 & 0.62 & 0.64 \\
{\bf Ours} & {\bf 0.72} & {\bf 0.64} & {\bf 0.68} & {\bf 0.62} & {\bf 0.72} \\
\bottomrule
\end{tabular}}
\vspace{-5pt}
\caption{Question quality ranking agreement between various methods and each evaluator (T1 through T5). Our metric achieves the best agreement with every evaluator. 
}
\label{tab:quality}
\end{table}

\begin{figure}[t]
    \centering
    \vspace{-45pt}
    \includegraphics[width=1.1\linewidth]{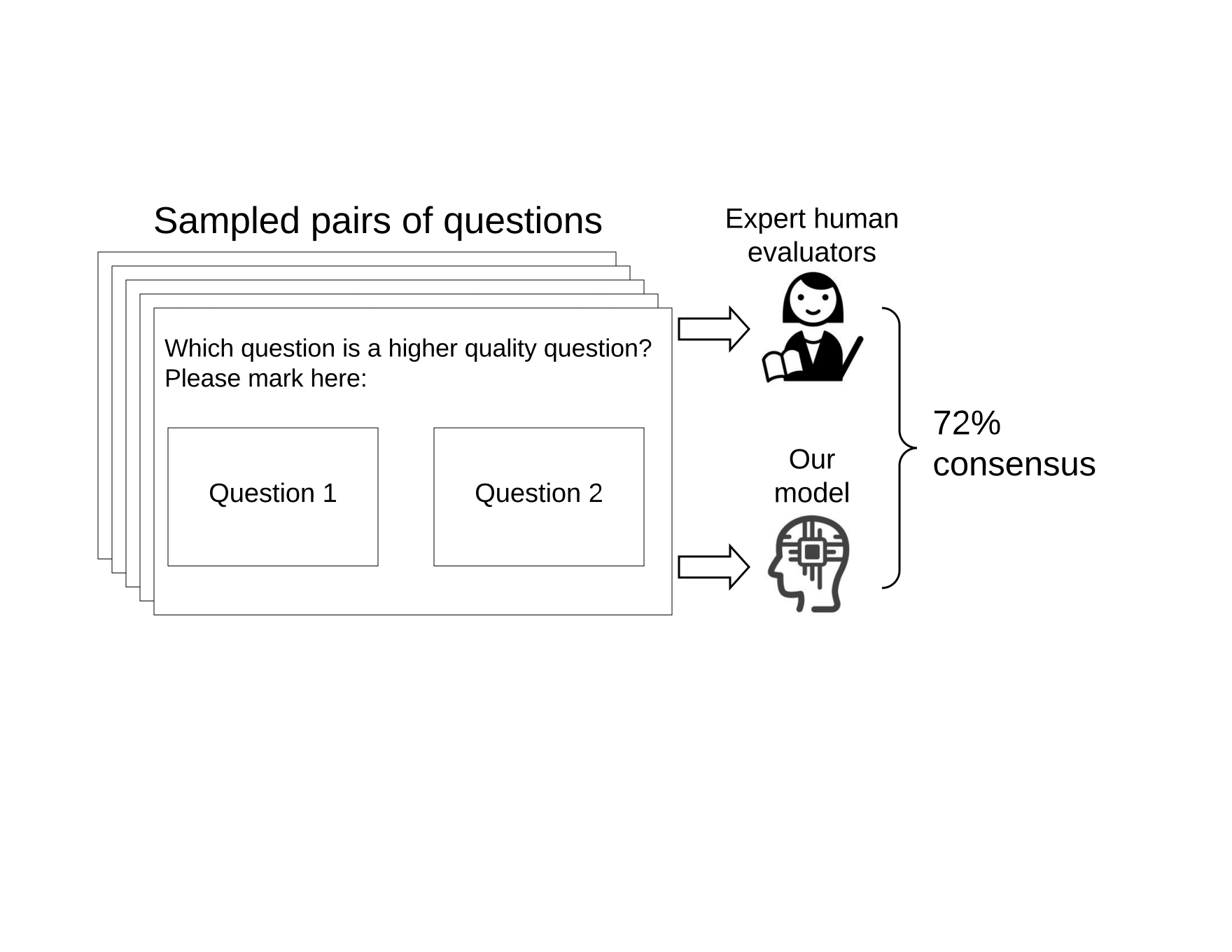}
    \vspace{-80pt}
    \caption{Illustration of question quality evaluation interface for the human evaluator. Our quality metric achieves a maximum of 72\% agreement with human evaluators.}
    \label{fig:question-quality-compare-human-machine}
\end{figure}

\paragraph{Results} Table~\ref{tab:spearmanr-scheme} shows the Spearman correlation coefficients comparing expert's topic and scheme rankings, respectively, to our framework's and two other baselines' rankings. The baselines include random ordering, using majority imputation to fill the data matrix, and using the observed data alone. We see that our framework's ranking closely matches the human expert's ranking while baselines do not produce rankings that are any close to the expert's ranking.

\begin{figure*}[t!]
\centering
\raisebox{0.15\height}{\includegraphics[width=0.24\linewidth]{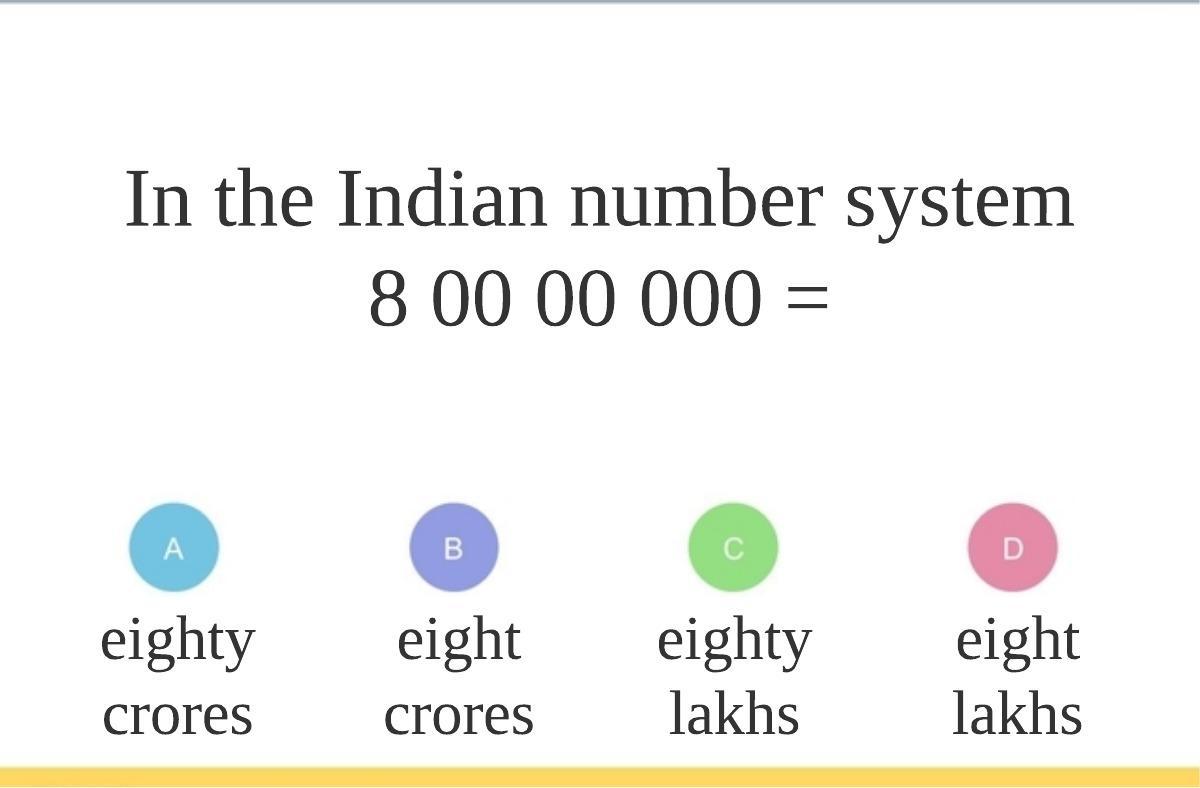}}
\hspace{-15pt}
\includegraphics[width=0.26\linewidth]{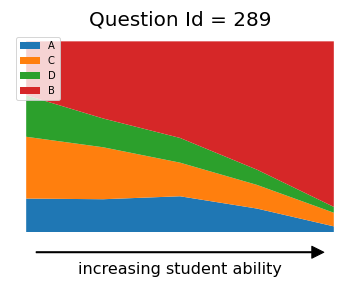}
\hspace{5pt}
\raisebox{0.15\height}{\includegraphics[width=0.24\linewidth]{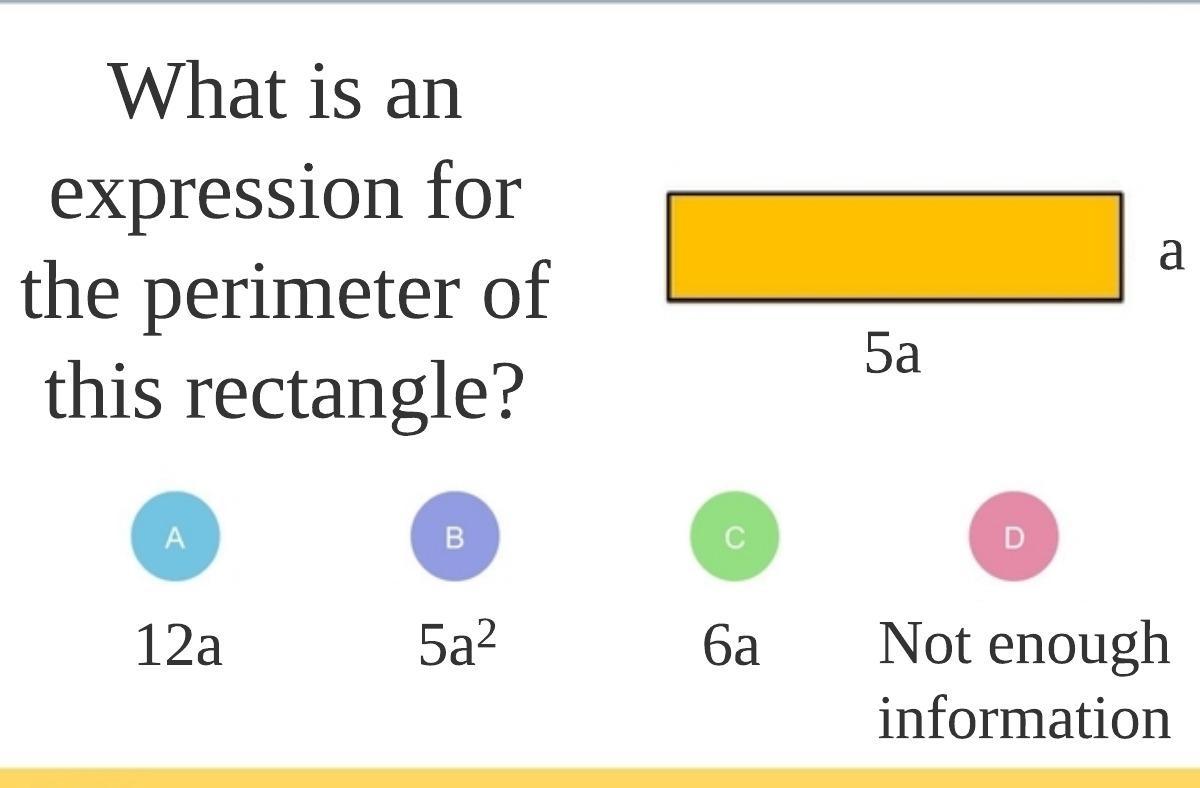}}
\hspace{-5pt}
\includegraphics[width=0.26\linewidth]{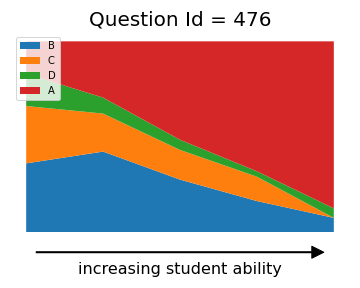}\\[1em]
\vspace{5pt}
\raisebox{0.15\height}{\includegraphics[width=0.24\linewidth]{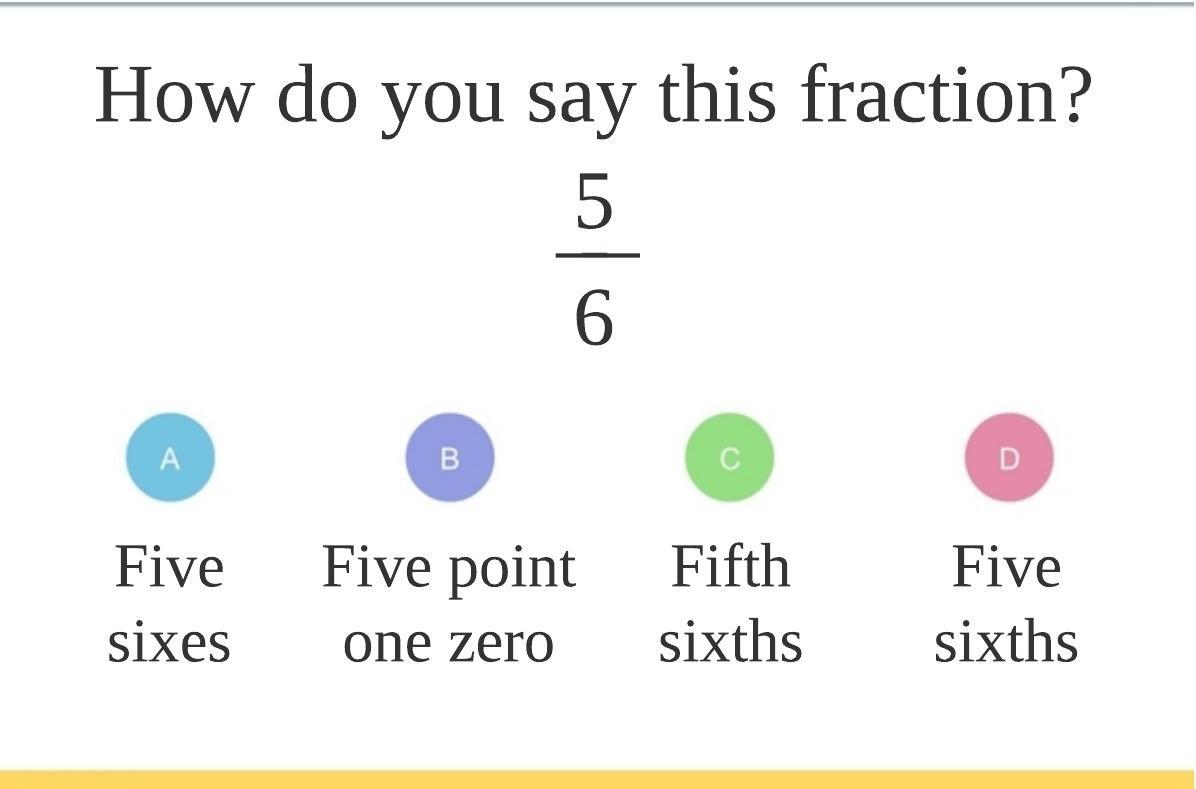}}
\hspace{-15pt}
\includegraphics[width=0.26\linewidth]{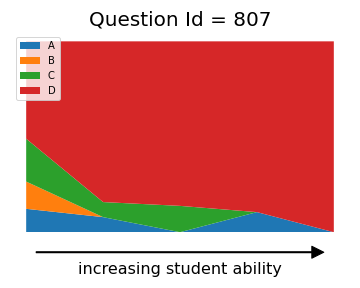}
\hspace{5pt}
\raisebox{0.15\height}{\includegraphics[width=0.24\linewidth]{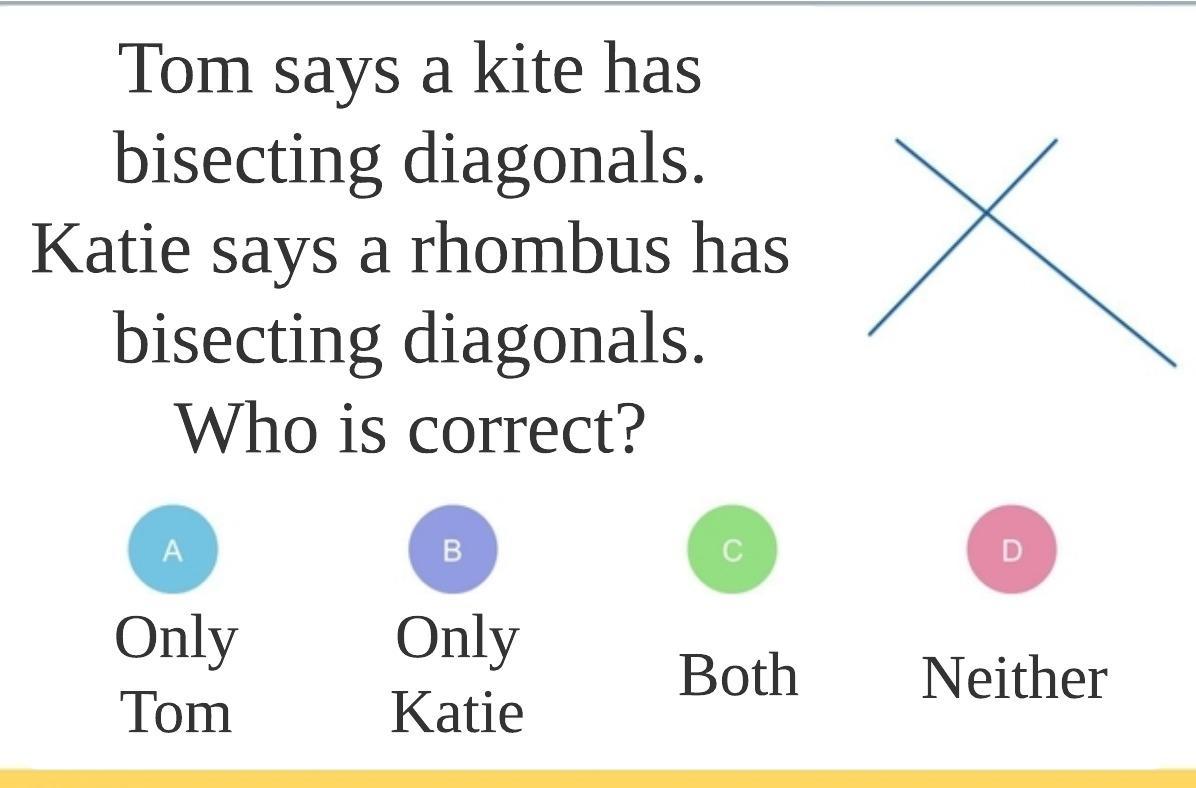}}
\hspace{-5pt}
\includegraphics[width=0.26\linewidth]{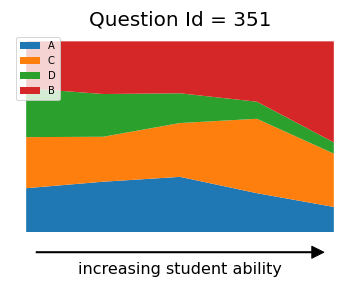}
\hspace{-5pt}
\caption{Two examples of high-quality questions (top row) and two examples of low-quality questions (bottom row) determined by our framework. For each pair, the left image shows the actual question, and the right image shows the stacked portion plot indicating the percentage of students who answered A, B, C, or D, where the top portion (red) always indicates the percentage of students who answered correctly. {\bf Top row}: high-quality questions differentiate students' abilities. {\bf Bottom left}: this low-quality question is too easy; students tend to answer it correctly despite their ability. {\bf Bottom right}: these low-quality questions are either too easy or too difficult; the majority of the students tend to answer it either correctly or incorrectly despite their ability.
}
\label{fig:question-quality-example-main-text}
\end{figure*}

\subsection{Question Quality Quantification}

\begin{figure}
    \centering
    \includegraphics[width=0.7\linewidth]{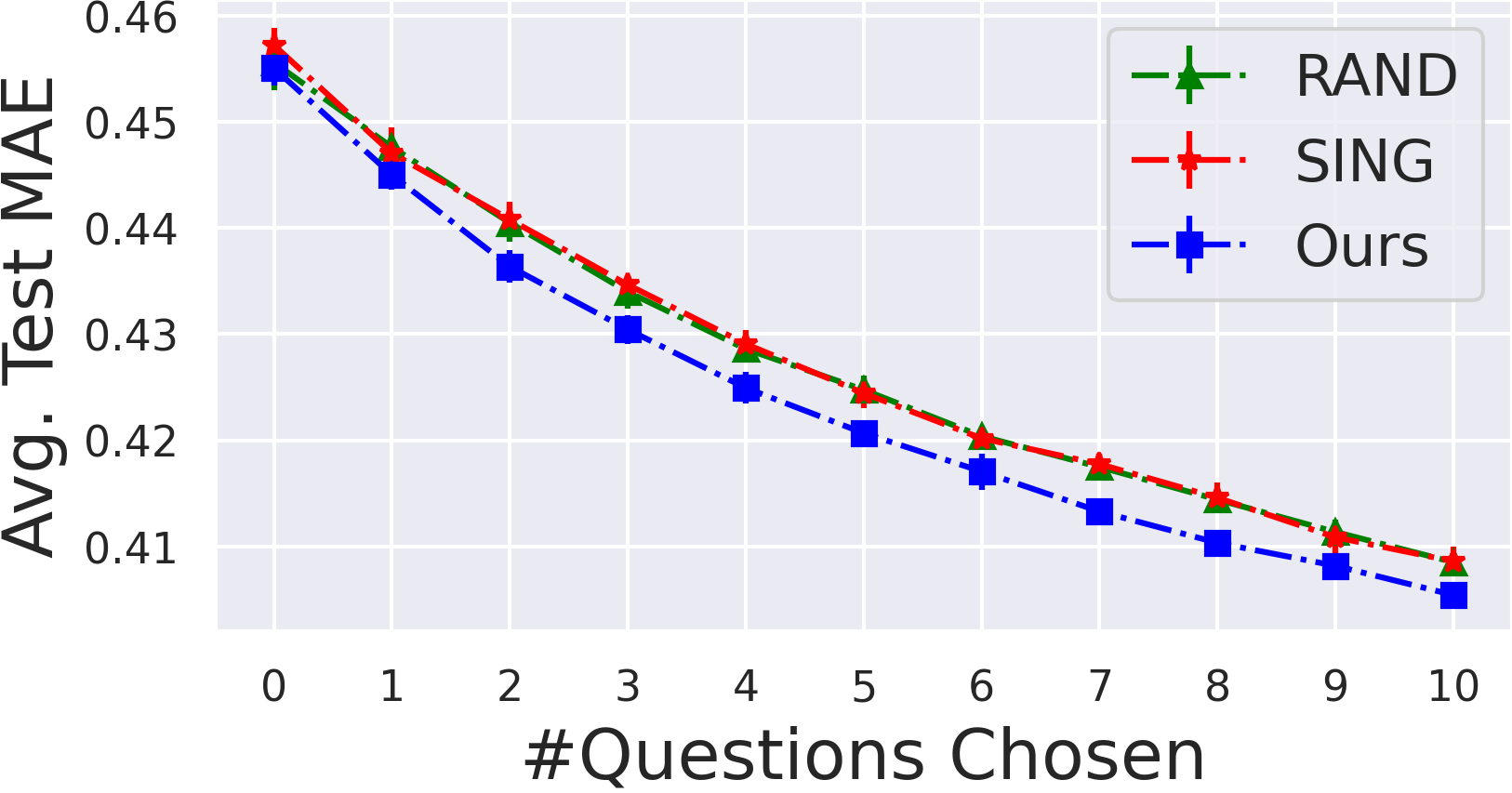}
    \caption{ Our personalized question selection strategy outperforms baseline strategies for each question selected.}
    \label{fig:active}
\end{figure}
\paragraph{Setup}
We compute question quality according to Eq.~\ref{eq:question-quality-form-2} and compare the pairwise rankings of question quality to those provided by human evaluators. 
Our 5 evaluators are all highly 
respected math teachers who have no prior information about this work.  
We resort to pairwise comparison, i.e., we give all 5 evaluators a pair of questions and ask them to give a preference on which question is of higher quality. We then compute the number of times our model agrees with each evaluator's choice of quality. Because this evaluation requires significant domain knowledge and because there are more than ten thousand questions and thus many more possible pairs, we only sample 80 pairs from a subset of our full dataset. Although limited, the evaluation of these samples provides preliminary evidence on whether the proposed quality metric agrees with domain experts. Figure~\ref{fig:question-quality-compare-human-machine} illustrates the interface that the evaluator sees when they provide the question quality labels. 
Each evaluator performed this task in isolation without knowledge of the other evaluators' labels.

We also consider two baseline metrics for question quality. The first metric randomly assigns a ranking to each question. The second metric computes the entropy of a question where the probability is the portion of correct or incorrect answers among all answer records for this question.

\begin{table*}
\centering
\vspace{-5pt}
{\small
\begin{tabular}{@{}lll@{}}
\toprule
{\bf Questions} & {\bf Student \#1} & {\bf Student \#2} \\ \midrule
Q1 & Squares, Cubes, etc. & Perimeter and Area \\
Q2 & Simplifying Expressions by Collecting Like Terms & Line Symmetry \\
Q3 & Squares, Cubes, etc & Missing Lengths \\
Q4 & Mental Multiplication and Division & Measuring Angles \\
Q5 & Mental Multiplication and Division & Line Symmetry\\
Q6 & Factors and Highest Common Factor & Basic Angel Facts \\
Q7 & Factors and Highest Common Factor & Angles\\
Q8 & Factors and Highest Common Factor & Measuring Angles \\
Q9 & Simplifying Expressions by Collecting Like Terms & Perimeter and Area \\
Q10 & Simplifying Expressions by Collecting Like Terms & Transformations \\
\bottomrule
\end{tabular}}
\caption{Example of the topics of the 10 questions that our framework chooses for 2 students. 
}
\label{tab:personalization-example}
\end{table*}

\paragraph{Results}
Table~\ref{tab:quality} shows the agreement between the models' question quality rankings and those provided by each of the 5 evaluators. We can see that our proposed quality metric achieves the highest agreement with human judgment. Also note that the entropy baseline, which is also inspired by information theory, although outperformed by our metric, is significantly better than random rankings. This preliminary result implies that further development of better question quality metrics via similar information-theoretic approaches has the potential to accurately capture human judgment of question quality and is a promising direction to pursue.

To visualize the high and low-quality questions that our framework selects, we show in
Figure~\ref{fig:question-quality-example-main-text} two examples of high-quality questions (top row) and two examples of low-quality questions (bottom row) determined by our framework. For each pair, the left image shows the actual question, and the right image shows the stacked portion plot. The stacked portion plot shows the percentage of students in different ability ranges
who have answered the question correctly. (the correct answer choice is always at the top, i.e., the red color part of the plot, and the remaining colors are the remaining three incorrect answer choices). The stacked portion plot is produced using the observed students' answer choices to the questions (i.e., A, B, C, or D choices). 

In addition to the question content itself, we can gain some insights by examining and comparing the stacked portion plots. For example, We can see that high-quality questions better test the variability in students' abilities because fewer students with a lower ability score can answer them correctly, whereas more students with a higher ability score can answer them correctly. This phenomenon is not present in lower quality questions, where most of the students, regardless of their ability score, tend to answer them either correctly or incorrectly. 

\vspace{-5pt}
\paragraph{Implications} Our difficulty and quality metrics efficiently and accurately provide data-driven analytics on questions which may help teachers in designing quizzes, homework sets, and exams that best suit their classes.

\subsection{Personalized Question Selection}
\paragraph{Setup}
To quantitatively evaluate the performance of our selection strategy, we proceed as follows.
We first sample a subset of students from the dataset and sample 10\% of each student's answer records as the prediction targets. For each student, we then sequentially choose 10 questions as prescribed in Eq.~\ref{eq:rec-1}, reveal their answers, and predict the student's answers using these 10 revealed answers through p-VAE. We report the average mean absolute error over all the sampled answer records at each of the 10 steps in the question selection process. We consider two baselines, a random selection strategy (RAND) and a single global optimal strategy (SING) which is similar to our proposed strategy but the reward is averaged over all students. Therefore, this strategy only selects one single sequence of questions for all students and cannot achieve personalization.

\paragraph{Results}
Figure~\ref{fig:active} reports the average MAE comparing our proposed strategy with RAND and SING after 10 runs. Our strategy achieves lower MAE at every step of the question selection process, demonstrating its effectiveness in choosing a sequence of questions that minimizes prediction error. 

\begin{figure}
    \centering
    \includegraphics[width=1\linewidth]{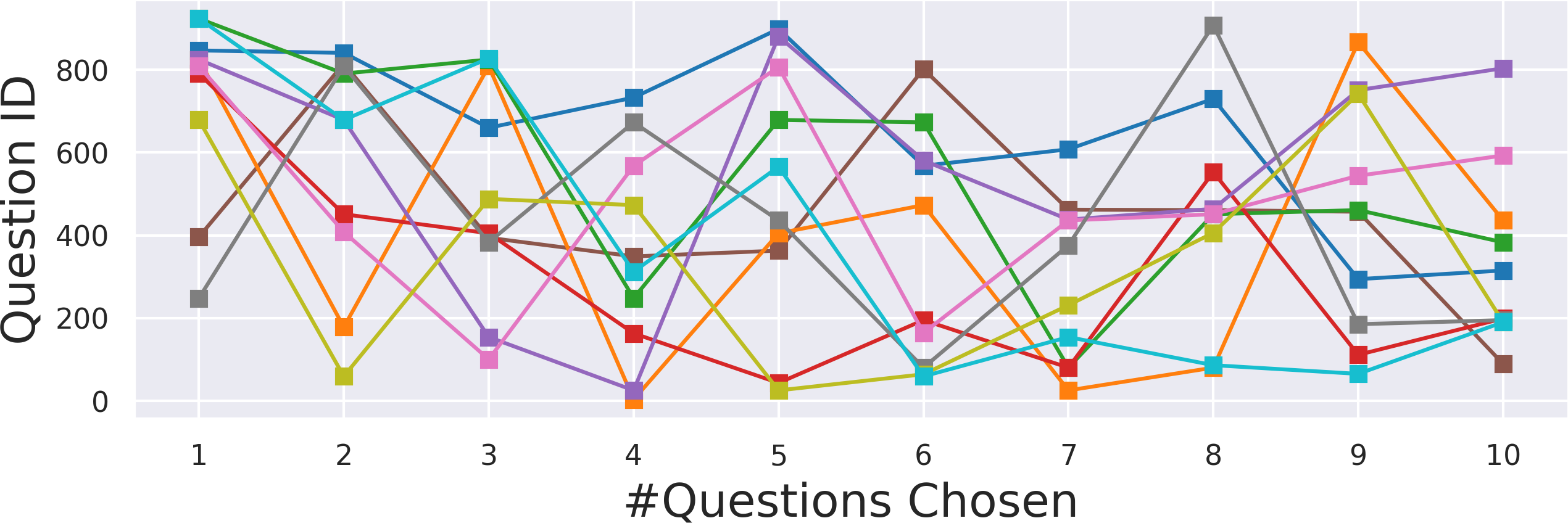}
    \caption{Illustration of the questions that our framework selects for 10 students. Each student gets a personalized sequence of questions.}
    \label{fig:personalization}
\end{figure}

To demonstrate that our strategy is personalized for each student, we show in figure~\ref{fig:personalization} the sequence of questions for 10 students. The x-axis is the step (number) of the question selected and the y-axis is the question ID. Each color represents the selected question sequence for one student. We can see that our strategy picks a very different set of questions for these 10 students. We also show the topics of the selected questions for two students in Table~\ref{tab:personalization-example} to gain some understanding of how our strategy chooses questions. We can see that the questions chosen for these two students are very different: the questions for the first students emphasize a few recurring topics while the questions for the second students cover a diverse set of topics. 

\paragraph{Implications} Our question selection strategy provides an efficient and automatic method to sequentially select questions for students. This can be used in large-scale learning and adaptive testing scenarios in which teachers cannot attend to each individual students but still would like to have a quick way to assess each student's skills.

%% file: related.tex
\section{Related Work}

\paragraph{AI in Education}
There is a vast and rapidly expanding literature on using AI for advancing education.
A sample of examples includes knowledge tracing~\cite{JMLR:v15:lan14a,rasch1960studies,Vie_Kashima_2019,Lan:2014:TLC:2623330.2623631,AAAI1817093,NIPS2015_5654}; 
various automation including grading~\cite{Waters:2015:BBA:2724660.2724672,Lan:2015:MLP:2724660.2724664}, feedback generation~\cite{Wu_Mosse_Goodman_Piech_2019} and quiz question generation~\cite{Wang:2018:QDQ:3231644.3231654}; and understanding students' online and offline behaviors such as collaboration~\cite{waters2014collaboration} and cheating~\cite{levitt2015catching}.

Most related to our work is prior literature on acquiring educational insights by analyzing students' answer records to questions. Notable examples include the Rasch model~\cite{rasch1960studies} that outputs a scalar question difficulty and a scalar student ability; the sparse factor analysis (SPARFA) model~\cite{JMLR:v15:lan14a} which extends the Rasch model to output multi-dimensional question and student analytics; the time-varying SPARFA model~\cite{Lan:2014:TLC:2623330.2623631} capable of processing time-varying students' answer records which is a more realistic scenario. These models, however, are limited to their high computational complexity and thus are difficult to apply to large scale educational data analysis. 
Our work complements prior work in that we develop a framework capable of efficiently analyzing large educational data sets while at the same time extracting educationally meaningful insights. 

\paragraph{Missing Data Imputation}
The missing Data Imputation method is used in many real-world applications, such as recommender systems.
Many existing approaches for recommender systems rely on linear methods because of their efficiency and scalability~\cite{Salakhutdinov:2007:PMF:2981562.2981720,8068603,stern2009matchbox}. More recent literature introduces nonlinear models using deep learning for improved model capacity, notably with variational autoencoders~\cite{Chen:2018:CVA:3270323.3270326,karamanolakis2018item,Li:2017:CVA:3097983.3098077,2018arXiv180703653N,Wang:2015:RSD:2888116.2888141}. However, many of these models do not handle missing data without some ad-hoc modifications to the data, i.e., zero imputation as in~\cite{2017arXiv170801715K,2018arXiv180205814L,sedhain2015autorec}. Such an ad-hoc way of replacing missing values in the data matrix significantly changes the original data distribution and can negatively impact the imputation results. 
In contrast to the above methods, p-VAE leverages amortized Bayesian inference with a special model architecture to efficiently handle missing data and quantify uncertainty, which is ideal for our application. 

Our work builds on~\cite{ma2018partial,eddi,gong2019icebreaker,mattei2018missiwae}, all of which take advantage of the desirable properties of p-VAE for recommender systems and information acquisition framework, respectively. Our work further advances such prior work in that we extend, to the first of our knowledge, p-VAE to the application of education data mining with a novel information criterion to extract educationally meaningful insights.

%% file: discussion.tex
\section{Conclusions}
In this paper, we develop a framework to analyze questions in online education platforms on a large scale. Our framework combines the recently proposed partial variational auto-encoder (p-VAE) for efficiently processing large scale, partially observed educational datasets, and novel metrics and strategy for automatically producing a suite of meaningful and actionable insights about quiz questions. We demonstrate the applicability of our framework on a  real-world educational dataset, showcasing the rich and interpretable information including question difficulty, question quality, and personalization that our framework obtains from millions of students' answer records to multiple-choice questions. 

Our framework is highly flexible, which enables further improvements and extensions to obtain richer educational insights. 
For example, one extension is to customize the information-theoretic metrics for extracting various other information of interest.
Another extension is to adapt the p-VAE model for time-series data, where we can work with a more realistic yet challenging scenario that students' states of knowledge change over time.  To facilitate and encourage future research, we have open sourced our dataset in the form of a competition for AI in education; see {{\url{https://eedi.com/projects/neurips-education-challenge}}} for more details.